\documentclass[preprint,12pt]{article}

\usepackage{epsfig}
\usepackage{amssymb}

\def\C {\chi_c}
\def\E {\eta_c}

\def\L {{\cal L}}
\def\M {{\cal M}}
\def\P {{\cal P}}
\def\R {{\cal R}}
\def\cc {$c\bar{c}$~}

\def\ap#1#2#3   {{\rm Ann. Phys. (NY)}       #1 (#3) #2}
\def\apj#1#2#3  {{\rm Astrophys. J.}         #1 (#3) #2}
\def\apjl#1#2#3 {{\rm Astrophys. J. Lett.}   #1 (#3) #2}
\def\app#1#2#3  {{\rm Acta. Phys. Pol.}      #1 (#3) #2}
\def\cpc#1#2#3  {{\rm Computer Phys. Comm.}  #1 (#3) #2}
\def\dum#1#2#3  {{~}                         #1 (#3) #2}
\def\epjc#1#2#3 {{\rm Eur. Phys. J. C}       #1 (#3) #2}
\def\err#1#2#3  {{\it Erratum}               #1 (#3) #2}
\def\ib#1#2#3   {{\it ibid.}                 #1 (#3) #2}
\def\jcp#1#2#3  {{\rm J. Comp. Phys.}        #1 (#3) #2}
\def\jmp#1#2#3  {{\rm J. Math. Phys.}        #1 (#3) #2}
\def\ijmp#1#2#3 {{\rm Int. J. Mod. Phys.}    #1 (#3) #2}
\def\jpg#1#2#3  {{\rm J. Phys. G.}           #1 (#3) #2}
\def\mpl#1#2#3  {{\rm Mod. Phys. Lett.}      #1 (#3) #2}
\def\nat#1#2#3  {{\rm Nature (London)}       #1 (#3) #2}
\def\ncim#1#2#3 {{\rm Nuovo Cimento}         #1 (#3) #2}
\def\nim#1#2#3  {{\rm Nucl. Instr. Meth.}    #1 (#3) #2}
\def\np#1#2#3   {{\rm Nucl. Phys.}           #1 (#3) #2}
\def\npb#1#2#3  {{\rm Nucl. Phys. B}         #1 (#3) #2}
\def\pan#1#2#3  {{\rm Phys. At. Nuclei}      #1 (#3) #2}
\def\pl#1#2#3   {{\rm Phys. Lett.}           #1 (#3) #2}
\def\plb#1#2#3  {{\rm Phys. Lett. B}         #1 (#3) #2}
\def\prep#1#2#3 {{\rm Phys. Rep.}            #1 (#3) #2}
\def\prev#1#2#3 {{\rm Phys. Rev.}            #1 (#3) #2}
\def\prc#1#2#3  {{\rm Phys. Rev. C}          #1 (#3) #2}
\def\prd#1#2#3  {{\rm Phys. Rev. D}          #1 (#3) #2}
\def\prev#1#2#3 {{\rm Phys. Rev.}            #1 (#3) #2}
\def\prl#1#2#3  {{\rm Phys. Rev. Lett.}      #1 (#3) #2}
\def\prs#1#2#3  {{\rm Proc. Roy. Soc.}       #1 (#3) #2}
\def\ptp#1#2#3  {{\rm Prog. Theor. Phys.}    #1 (#3) #2}
\def\ps#1#2#3   {{\rm Physica Scripta}       #1 (#3) #2}
\def\rmp#1#2#3  {{\rm Rev. Mod. Phys.}       #1 (#3) #2}
\def\rpp#1#2#3  {{\rm Rep. Prog. Phys.}      #1 (#3) #2}
\def\sjnp#1#2#3 {{\rm Sov. J. Nucl. Phys.}   #1 (#3) #2}
\def\spj#1#2#3  {{\rm Sov. Phys. JETP}       #1 (#3) #2}
\def\spu#1#2#3  {{\rm Sov. Phys.-Usp.}       #1 (#3) #2}
\def\yaf#1#2#3  {{\rm Yad. Fiz.}             #1 (#3) #2}
\def\zp#1#2#3   {{\rm Zeit. Phys.}           #1 (#3) #2}
\def\zpa#1#2#3  {{\rm Zeit. Phys. A}         #1 (#3) #2}
\def\zpc#1#2#3  {{\rm Zeit. Phys. C}         #1 (#3) #2}
\def\et{{\rm et al.,}}
\begin{document}

\begin{center}
{\Huge Touching on the gluon polarization}\\ 
{\Huge in the Durham Pomeron}\\[5mm]

S.P.\ Baranov\\[5mm]

{\sl P.N. Lebedev Institute of Physics, Moscow, Russia}\\
{e-mail: baranov@sci.lebedev.ru}\\[5mm]
\end{center}

\parbox{12cm}{{\small
Inclusive and exclusive production of scalar and pseudoscalar mesons 
via gluon-gluon fusion is considered.
A new experimental test is proposed, which can probe the polarization 
state of the gluons in exclusive production processes and check
the compatibility of the two-gluon exchange mechanism with the
properties of conventional Pomeron.\\[5mm]
}}\\

\section{Introduction}

According to the present-day point of view, diffractive processes are
to be understood in terms of two-gluon exchange mechanism, and the old 
concept of Pomeron has to be replaced with a new field theory object,
the generalized two-gluon distribution function. It is not yet clear, 
however, to what extent does the two-gluon language fit the known 
properties of conventional Pomeron.
In particular, since the Pomeron is considered to be a scalar object 
with positive parity, $J^{PC}=0^{++}$, the polarization vectors of the 
both gluons must be parallel to each other. Do we have any chance to 
verify this prediction experimentally? The present Letter is devoted 
to this interesting issue, and, as far as the Durham model \cite{KhMR}%
-\cite{Szcz} is concerned, the answer is yes, as we will see in the next 
section.

\section{Basic idea}

In order to feel the correlation between the gluon polarization vectors,
we have to compare the production cross sections of pseudoscalar and 
scalar mesons in inclusive and exclusive channels. Feynman diagrams 
representing the inclusive inelastic and exclusive double diffractive
production mechanisms are displayed in Fig. 1(a) and (b), respectively.
Both these mechanisms have an important element in common, namely, the
gluon-gluon fusion partonic subprocess. The properties of the 
corresponding matrix element $\M$ are such that, if the polarization 
vectors of the incident gluons are parallel to each other, then only
the scalar states are produced, while the production of pseudoscalar
states is forbidden. At the same time, if the polarization vectors of 
the initial gluons are perpendicular to each other, then the production
of pseudoscalar states is allowed, while the production of scalar states 
is suppressed.

These properties follow immediately from the interaction Lagrangians,
which read (for an ideal case of on-shell gluons)
\begin{equation}\label{plus}
\L(gg0^{+}) \propto e_1^\mu e_2^\nu g_{\mu\nu}
\end{equation}
for scalar states, and
\begin{equation}\label{minus}
\L(gg0^{-}) \propto \epsilon_{\mu\nu\alpha\beta}
 e_1^\mu e_2^\nu k_1^\alpha k_2^\beta
\end{equation}
for pseudoscalar states, with $e_1$ and $e_2$ being the gluon 
polarization vectors, and $k_1$ and $k_2$ the gluon momenta.

In the inelastic case, the gluons $g_1$ and $g_2$ in Fig. 1(a) are
uncorrelated and contribute to the production of both pseudoscalar and
scalar mesons. On the contrary, in the diffractive case, the presence
of an additional gluon $g_0$ in Fig. 1(b) can induce positive 
correlations. If we assume that the emission of a gluon pair $g_0g_1$
is equivalent to the emission of a Pomeron, then the gluons $g_0$ and 
$g_1$ are polarized in the same plane. For the same reason, the gluons
$g_0$ and $g_2$ are also polarized in the same plane, and so, the 
gluons $g_1$ and $g_2$ have to be polarized in the same plane as well.
Consequently, one can expect that the production of pseudoscalar mesons 
would be suppressed or forbidden in this case.

Thus, by comparing the pseudoscalar-to-scalar ratios in exclusive and 
inclusive production channels one probes the polarization of the 
interacting gluons. Though evident, the suppression of $J^P=0^-$ states 
due to quantum number selection rules has not been taken into account 
in the literature \cite{KhMR}-\cite{Szcz} when considering the elastic 
production of non-standard Higgs bosons, $\eta'$ and $\eta_c$ mesons.

\section{Numerical example}

To illustrate the analysing power of the gluon-gluon fusion subprocess
we show theoretical predictions on the inclusive production of 
$\E\,(J^{PC}=0^{-+})$ and $\C\,(J^{PC}=0^{++})$ mesons at the conditions 
of the Fermilab Tevatron ($p\bar{p}$ collisions at $\sqrt{s}=1960$ GeV) 
and Relativistic Heavy Ion Collider (RHIC, $pp$ mode, $\sqrt{s}=200$ GeV).

Throughout this Letter we will be using the overall (proton-proton)
center-of-mass system with the $z$ axis oriented along the beam direction.
Adopting the $k_t$-factorization prescription \cite{GLR}, we take the
gluon polarization vectors in the form $e^\mu = k_t^\mu/|k_t|$, where
$k_t^\mu$ is the component of the respective gluon momentum perpendicular
to the beam axis. In this approach, the polarization vectors are real
and lie entirely in the $xy$ plane.

Our calculations are based on perturbative QCD and nonrelativistic
bound state formalism \cite{BaiBer,GubKra}. The creation of a heavy 
quark pair \cc is treated as a purely perturbative process.
The calculations are straightforward and follow the standard Feynman
rules. The spin projection operators \cite{BaiBer,GubKra}
\begin{eqnarray}
\label{j00} \P(^1S_0)\equiv
\P(S{=}0,L{=}0)&=&\gamma_5\,(\not{p_c}+m_c)/(2m_c)^{1/2} \\
\label{j11} \P(^3P_0)\equiv
\P(S{=}1,L{=}1)&=&
(\not{p_{\bar{c}}}-m_c)\not{\epsilon_S}\,(\not{p_c}+m_c)/(2m_c)^{3/2}
\end{eqnarray}
introduced in the amplitudes guarantee the proper quantum numbers of the 
\cc states under consideration.

The formation of a final state meson is a non-perturbative process.
Within the nonrelativistic approximation which we are using, its 
probability is determined by a single parameter related to the meson 
wave function at the origin of coordinate space. 
More precisely, the probability of forming an $\E$ meson is proportional 
to the radial wave function squared, $|\R_{\eta}(0)|^2$; the latter is 
taken to be the same as that of the $J/\psi$ meson (known from the 
$J/\psi$ leptonic decay width \cite{PDG}) and set to
$|\R_{\eta}(0)|^2 = 0.8$ GeV$^3$. 
The probability of forming a $\C$ meson relates to the derivative of
the radial wave function; the latter is taken from the potential model
\cite{EicQui}: $|\R'_{\chi}(0)|^2 = 0.075$ GeV$^5$.

Finally, we obtain the meson production cross section by convoluting the 
matrix element squared with unintegrated gluon distribution functions. 
Here, we use the parametrization proposed by J.Bl\"umlein \cite{Bluem},
where the leading-order Gl\"uck-Reya-Vogt gluon density \cite{GRV98}
was used for collinear input. The details of calculations from the first 
to the last step are explained in Ref. \cite{j_sha}. The full 
{\sc fortran} code is available from the author on request.

Fig. 2 displays our predictions on the transverse momentum spectra.
All of the pseudoscalar $\E$ mesons come from gluons with
perpendicular polarization vectors. Most of the scalar $\C$ mesons
come from gluons with parallel polarization vectors. As a result
of the initial gluon off-shellness, the production of $\C$ mesons by
gluons with perpendicular polarization vectors is not totally forbidden
(as it would follow from (\ref{plus})), but is only suppressed by
a large factor. This contribution comes from the interaction of the type
$\L(gg0^{+}) \propto (e_1 k_2)(e_2 k_1)$ which vanishes in the on-shell
gluon limit.

The ratio of the production rates $d\sigma(\E)/d\sigma(\C)$ is exhibited
in Fig. 3. As one can see, it stays almost constant at about
$d\sigma(\E)/d\sigma(\C)\simeq 5$ both at the Tevatron and RHIC energies,
neither showing dependence on the transverse momentum nor on rapidity.
Any deviation from this number would indicate correlations between the
gluon polarization states.

It is worth noting that many important theoretical uncertainties cancel
out in the ratio: e.g., the predictions are insensitive to the choice of
gluon distribution functions, the scale in the running coupling constant
$\alpha_s(\mu^2)$, etc. Thus, we conclude that the gluon-gluon fusion
mechanism can serve as a convenient tool probing the gluon polarization.

\section{Comparison with other calculations}

The properties of the gluon-gluon-meson vertex has been previously 
considered in Refs. \cite{Durham,KKhMR}. We fully agree with the mentioned 
papers in the general structure of this coupling (Eqs. (1) and (2) versus
Eq. (22) in Ref. \cite{Durham}). In the inelastic case, the angle between
the gluon polarization vectors is at the same time the azimuthal angle 
between the recoil protons. This results in the $\cos^2\varphi$ and
$\sin^2\varphi$ behavior of the production cross section for $0^+$ and
$0^-$ mesons, respectively. 
The behavior of double diffractive processes is less simple and strongly 
depends on the dynamics. Particularly essential is the relation between 
the internal transverse momentum $Q_T$ in the loop (see Fig. 3 of Ref.
\cite{Durham}) and the transverse momenta of the outgoing protons $p_t$.
The distribution of $0^+$ mesons preserves its $\cos^2\varphi$ behavior
for $Q_T<<p_t$ (which is typical for comparatively light states) and
becomes nearly flat for $Q_T>>p_t$ (for heavier states like Higgs bosons).

In the present Letter, we basically focus on the quantum number selection 
rules and avoid giving numerical predictions on the double diffractive
processes, as they would be too much model dependent. First of all, the 
very factorization principle is questionable. In order that the 
factorization be valid, both gluons attached to the `upper' proton in
Fig. 1 must only have positive light-cone momentum fraction, while the
negative one must be close to zero. Also, both gluons attached to the
`lower' proton must only have negative light-cone momentum fraction, 
while the positive one must be close to zero. These two conditions are 
incompatible for the gluon connecting the both protons. A possible way
out is in assuming that the positive and negative light-cone momenta
are both close to zero, and only the transverse momentum is exchanged.
But then it would be hard to believe that there exists a simple relation
between the generalized two-gluon distribution functions and the ordinary
gluon distributions measured in inclusive processes.

Instead, we give the first priority to clarifying the fundamental role
of selection rules coming from gluon spin correlations, that may have
dramatic effect on the production of all pseudoscalar states (including
non-standard Higgs bosons) but yet had never been considered in the
literature.

The first experimental observation of the double diffractive production
of scalar $\chi_{c0}$ mesons is recently reported in \cite{CDF}. This is
an encouraging result showing that the test which we propose here is 
really feasible.


\section{Summary}

In the present Letter we draw attention to the fact that measuring
the production cross sections of pseudoscalar and scalar mesons in 
inclusive and exclusive channels can yield important information on
the polarization state of the interacting gluons. 
In addition, it would be extremely useful if these measurements are
accompanied by measuring the azimuthal dependence of the cross section.

Whether or not will the Durham model receive support from the data,
the information obtained by these measurements will shed more light on 
the nature of diffractive processes and, in particular, will stimulate 
further refinement of the concept of generalized gluon distributions.

\section*{Acknowledgements}

This work was supported by DESY Directorate in the framework of
Moscow-DESY project on Monte-Carlo implementation for HERA-LHC,
by Russian Foundation for Basic Research under Grant No. 08-02-00896-a
and by FASI under Grant No. NS-1856.2008.2.

\newpage
\begin{figure}
\begin{center}
\epsfig{figure=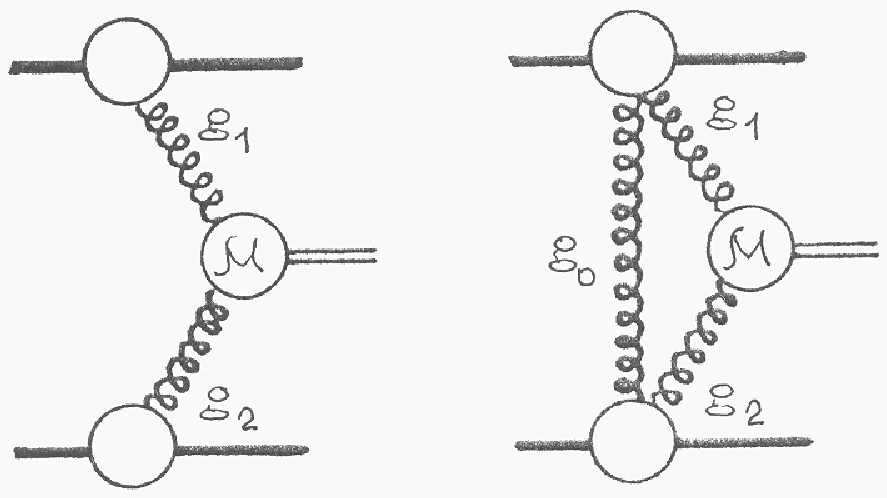,width=13cm}
\end{center}
\vspace*{-0.5cm}
\caption{Diagrams representing the inclusive (left panel) and
         exclusive (right panel) production mechanisms. $\M$ denotes 
         the gluon-gluon fusion matrix element; open circles stand for 
         the gluon distribution functions in the proton.}
\label{fig1}
\end{figure}

\begin{figure}
\begin{center}
\epsfig{figure=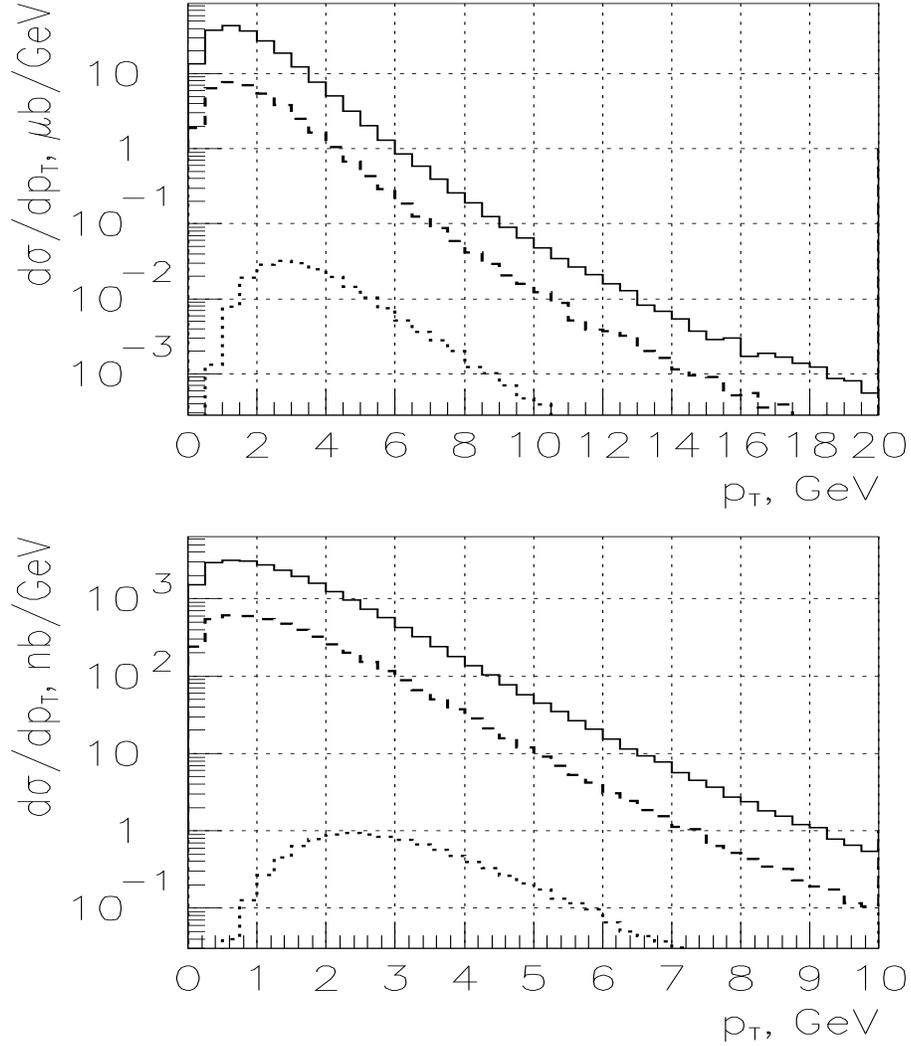,width=13 cm, height=15cm}
\end{center}
\vspace*{-0.5cm}
\caption{Inclusive transverse momentum distributions at the Tevatron
         (upper panel) and RHIC (lower panel) energies.
 Solid histograms, $\eta_c$ mesons, all coming from the gluons with
 perpendicular polarization vectors;
 dashed and dotted histograms, $\chi_c$ mesons coming from the gluons 
 with parallel and perpendicular polarization vectors, respectively.}
\label{fig2}
\end{figure}

\begin{figure}
\begin{center}
\epsfig{figure=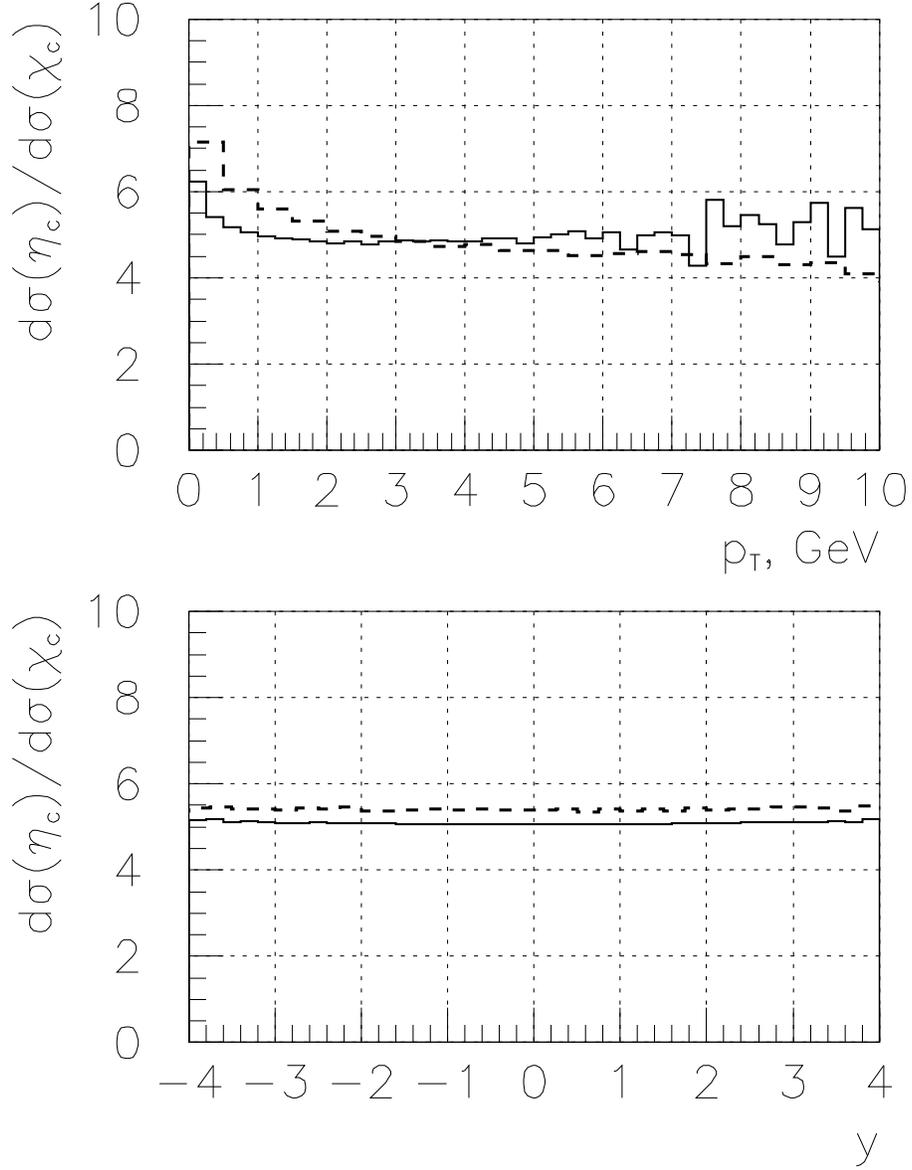,width=13cm}
\end{center}
\vspace*{-0.5cm}
\caption{The ratio of the production rates 
         $d\sigma(\eta_c)/d\sigma(\chi_c)$ as a function of the transverse 
         momentum $p_t$ (upper panel) and rapidity $y$ (lower panel).
 Solid histograms, RHIC conditions;
 dashed histograms, Tevatron conditions.}
\label{fig3}
\end{figure}
\end{document}